\documentclass[sn-mathphys-num]{sn-jnl}% Math and Physical Sciences Numbered Reference Style 
\usepackage{graphicx}%
\usepackage{multirow}%
\usepackage{amsmath,amssymb,amsfonts}%
\usepackage{amsthm}%
\usepackage{mathrsfs}%
\usepackage[title]{appendix}%
\usepackage{xcolor}%
\usepackage{textcomp}%
\usepackage{manyfoot}%
\usepackage{booktabs}%
\usepackage{algorithm}%
\usepackage{algorithmicx}%
\usepackage{algpseudocode}%
\usepackage{listings}%
\theoremstyle{thmstyleone}%
\theoremstyle{thmstyletwo}%

\theoremstyle{thmstylethree}%

\newcommand{\cl}{C \kern -0.1em \ell}
\begin{document}

\title[Article Title]{Hydrogen atom as a nonlinear oscillator under circularly polarized light: epicyclical electron orbits}


\author*[1]{\fnm{Quirino} \sur{Sugon Jr}}\email{qsugon@gmail.com}

\author[1]{\fnm{Clint Dominic G.} \sur{Bennett}}\email{cbennett@ateneo.edu}

\author[1,2]{\fnm{Daniel J.} \sur{McNamara}}\email{dmcnamara@observatory.ph}

\affil*[1]{\orgdiv{Department of Physics, School of Science and Engineering}, \orgname{Ateneo de Manila University}, \orgaddress{\street{Loyola Heights}, \city{Quezon City}, \postcode{1108}, \country{Philippines}}}

\affil[2]{\orgname{Manila Observatory}, \orgaddress{\street{Ateneo de Manila University Campus, Loyola Heights}, \city{Quezon City}, \postcode{1108}, \country{Philippines}}}


\abstract{We used Clifford algebra $Cl_{2,0}$ to find the 2D orbit of a hydrogen electron under a Coulomb force and a perturbing circularly polarized electric field of light at angular frequency~$\omega$, which is turned on at time $t = 0$ via a unit step switch. Using a coordinate system co-rotating with the electron's unperturbed circular orbit at angular frequency $\omega_0$, we derived a complex differential equation that is similar to but different from that of the Lorentz oscillator equation for light--atom interaction. We solved the homogeneous and particular differential equation and showed that the position of the electron is a linear combination of five exponential Fourier terms or orbital wave functions with  frequencies $0$, $\omega_0$, $2\omega_0$, $(2\omega_0 - \omega)$, and $\omega$, whose coefficients depend on the light-to-atom frequency ratio $\omega/\omega_0$ and light-to-atom force magnitude ratio $A/r_0$. We showed that the electron orbits are approximately Keplerian at light-to-atom frequency ratio $\alpha = \omega/\omega_0 = \{0, 1, 2\}$, with the orbits becoming discontinuous and divergent as $\alpha \rightarrow 1^{\pm}$, but continuous and non-divergent at $\alpha = \{0, 2\}$. These Keplerian orbits are approximated by the sum of the zeroth, first, and second harmonics of the electron's unperturbed orbital wave function $\hat\psi_0 = e^{\hat\imath\omega_0 t}$, corresponding to the eccentric, deferent, and epicycle in the Copernican construction of planetary orbits.}

\keywords{classical electrodynamics; light--atom interaction; circularly polarized light; nonlinear differential equation; exponential Fourier analysis; orbital mechanics; complex analysis; Clifford analysis}
\pacs[MSC Classification]{15A67, 34C15, 34M04, 42A16, 70F07, 70K60, 78A35, 78A40}

%15A67 Applications of Clifford algebras to physics, etc. 
%34C15 Nonlinear oscillations and coupled oscillators for ordinary differential equations 
%34M04 Nonlinear ordinary differential equations and systems in the complex domain
%42A16 Fourier coefficients, Fourier series of functions with special properties, special Fourier series
%70F07 Three-body problems
%70K60 General perturbation schemes for nonlinear problems in mechanics 
%78A35 Motion of charged particles
%78A40 Waves and radiation in optics and electromagnetic theory 



\maketitle
%\tableofcontents

\section{Summary of Changes}

Our original paper with the same title was published in ArXiv (2024)\cite{SugonBennettMcNamara2024}:
\begin{quote}
In this paper, we use Clifford algebra $\cl_{2,0}$ to find the 2D orbit of Hydrogen electron under a Coulomb force and a perturbing circularly polarized electric field of light at angular frequency~$\omega$, which is turned on at time $t = 0$ via a unit step switch. Using a coordinate system co-rotating with the electron's unperturbed circular orbit at angular frequency $\omega_0$, we derive the complex nonlinear differential equation for the perturbation which is similar to but different from the Lorentz oscillator equation: (1) the acceleration terms are similar, (2) the damping term coefficient is not real but imaginary due to Coriolis force, (3) the term similar to spring force is not positive but negative, (3) there is a complex conjugate of the perturbation term which has no Lorentz analog but which makes the equation nonlinear, and (4) the angular frequency of the forcing term is not $\omega$ but $\omega - \omega_0$. By imposing that the position and velocity of the electron are continuous at time $t = 0$, we show that the orbit of the electron is a sum of five exponential Fourier terms with frequencies 0, $\omega_0$, $2\omega_0$, $(2\omega_0 - \omega)$, and $\omega$, which correspond to the eccentric, deferent, and three epicycles in Copernican astronomy. We show that at the three resonant light frequencies $0$, $\omega_0$, and $2\omega_0$, the electron's orbit is divergent, but approximates a Keplerian ellipse. At other light frequencies, the orbits are nondivergent with periods that are integer multiples of $\pi/\omega_0$ depending on the frequency ratio $\omega/\omega_0$. And as $\omega/\omega_0\rightarrow \pm\infty$, the orbit approaches the electron's unperturbed circular orbit.
\end{quote}
But we made a mistake in this paper in the sign of the exponential: we wrote $\hat c_{-1} = -3e^{-2\phi_0\hat\imath}\hat c_1^*$, where $\phi_0$ is the orbital phase of the electron's unperturbed orbit just before the circularly polarized light is switched on. This equation should have been written as $\hat c_{-1} = -3e^{2\phi_0\hat\imath}\hat c_1^*$, which agrees with the Copernican eccentric--epicycle relation for planetary orbits\cite{SugonBragaisMcNamara2008}. Our conclusion then was that the electron orbits are all divergent at the three resonant light-to-atom frequency ratios: $\alpha = \omega/\omega_0 = \{0, 1, 2\}$. But this is only partially true.

We corrected these errors in our revised paper, which was published in the journal Hydrogen (2026)\cite{SugonBennettMcNamara2026}. In this revised paper, we showed that the orbit of the Hydrogen electron under circularly polarized light is discontinuous and divergent at $\alpha = 1$, but finite and quasi-Keplerian at light-to-atom frequency ratio $\alpha = \{0, 2\}$. In addition, we plotted the orbits instead of simply identifying parts of the exponential Fourier terms in terms of the Copernican notions of eccentric, deferent, and epicycle. Specifically, we plotted the orbits at different light-to-atom frequency ratios  subject to the phase condition $\phi_0 = \delta = 0$, where $\delta$ is the phase of the electric field $\mathbf E$ of the circularly polarized light. And for resonant and near-resonant light-to-atom frequency ratios, we plotted the orbits for $\phi_0 = 0$, but setting $\delta = \{0, \pi/2, \pi, 3\pi/2\}$. In all these plots, we kept the light-to-atom force magnitude ratio $A/r_0 \leq 0.2$ in order to satisfy the linear perturbation assumption.

\section{Significance}

The 2D interaction of a hydrogen atom and a circularly polarized light has no known closed-form solutions, because it belongs to a class of difficult problems called the restricted three-body problem. In our revised paper, we showed that if we apply first-order perturbation theory, we can solve the resulting nonlinear (antilinear) differential equation exactly using exponential Fourier series analysis. We did not use the standard techniques of Hamiltonian mechanics as done by most of the other authors, but instead used vectors and complex numbers in Newtonian Mechanics within the framework of Clifford (geometric) algebra~$Cl_{2,0}$. 

Unlike the Lorentz oscillator model that yields only one resonant frequency $\omega = \omega_0$, our oscillator model of the atom yielded three resonant frequencies: $\omega = \{1, \omega_0, 2\omega_0\}$. These frequencies correspond the three Copernican harmonics: eccentric, deferent, and epicycle. As $\omega \rightarrow \omega_0$, the orbit becomes discontinuous and divergent. On the other hand, as $\omega \rightarrow \{0, 2\omega_0\}$, the orbit becomes continuous and non-divergent, because the divergent terms in the $\hat c$ coefficients cancel out those in the $\hat b$ coefficients. 

We showed that for light-to-atom frequency ratios satisfying $\omega/\omega_0 \neq 1$, we can plot the new orbit of the hydrogen electron from its initial circular orbit when the light is switched on, since the $x$ and $y$ positions of the electron are explicitly expressed in terms of cosine and sine functions of light's electric field amplitude $|\hat a|$, angular frequency $\omega$, rotational phase $\delta$ and of the electron's unperturbed orbital radius $|\hat r_0|$, orbital angular frequency $\omega_0$, and orbital phase $\phi_0$. These analytical expressions for the plots of the electron orbits allow us to use simple plotting software (e.g., Google Sheets 2026-04-06 for the tables of orbit coefficients and Tikz 2023-01-15 v3.1.10 for the plots of orbits.) for different values of time $t$ instead of using complicated numerical algorithms that require the time step $\Delta t$ to be as small as possible.

Moreover, unlike the Bohr--Sommerfeld planetary model of the atom, we did not assume that the atom only absorbs radiation in quantized or discrete frequency values. Instead, we assumed that the atom can absorb radiation at any frequency. There are no quantum jumps in our model, i.e., the electron changes the radius of its orbit at infinitesimally small time. Instead, we assumed that the electron's position and velocity are continuous just  before and just after the circularly polarized light is switched on at time $t = 0$. It is only when the light is switched on when the electron changes its orbit from circular to a sum of different circular orbits at different frequencies. 

We expect that if we extend the perturbation theory from first-order to second-order and higher, we shall generate more exponential Fourier terms to describe the hydrogen electron orbits, whose frequencies are integral or fractional multiples of the fundamental angular frequency $\omega_0$ of the electron's unperturbed circular orbit. From these orbits we may be able to  compute the refractive index and energy absorption spectra of a hydrogen atomic gas, which can serve as a better model than the Lorentz model of light--atom interaction. This would require a careful revisiting of the classical radiation theory without invoking the Bohr--Sommerfeld quantization rules.

\section{Future Works}
  
In future studies, we shall extend our work on hydrogen electron orbits under circularly polarized light by including the magnetic field of light in the force equation and employing second-order perturbation theory to compute the exponential Fourier series coefficients and wave functions, in order to  compute the electrical permittivity, refractive index, and absorption spectra of the hydrogen atom.

\subsection*{Author Contributions}

Q.S.J. derived the theoretical equations, wrote the manuscript, computed the tables, and drew the plots. C.D.G.B. and D.J.M. reviewed the equations, tables, and plots; they also gave corrections, suggestions, and additional references. All authors have read and agreed to the published version of the manuscript.

\subsection*{Funding}
Q.S.J. was funded in this research by Ateneo de Manila University through the Loyola Schools Scholarly Work Grant (SOSE 04 2023) from 1 January to 31 December 2023. The authors wish to thank the University Research Council of Ateneo de Manila University for the open access publication support. 

\subsection*{Acknowledgments}

Q.S.J. used Wolfram Alpha online to simplify the $\alpha_1$ polynomials in the $\hat c$ and $\hat b$ coefficients. He also used the Tikz package in \LaTeX\ to make the final plots.



%%\begin{thebibliography}{100}

%%\end{thebibliography}
%\bibliography{paper-001-bibliography-20240118}% common bib file
%\bibliography{sugon-20240928-biblio-hydrogen-atom-nonlinear-oscillator-circularly-polarized-light}% common bib file

\begin{thebibliography}{60}
\ifx \bisbn   \undefined \def \bisbn  #1{ISBN #1}\fi
\ifx \binits  \undefined \def \binits#1{#1}\fi
\ifx \bauthor  \undefined \def \bauthor#1{#1}\fi
\ifx \batitle  \undefined \def \batitle#1{#1}\fi
\ifx \bjtitle  \undefined \def \bjtitle#1{#1}\fi
\ifx \bvolume  \undefined \def \bvolume#1{\textbf{#1}}\fi
\ifx \byear  \undefined \def \byear#1{#1}\fi
\ifx \bissue  \undefined \def \bissue#1{#1}\fi
\ifx \bfpage  \undefined \def \bfpage#1{#1}\fi
\ifx \blpage  \undefined \def \blpage #1{#1}\fi
\ifx \burl  \undefined \def \burl#1{\textsf{#1}}\fi
\ifx \doiurl  \undefined \def \doiurl#1{\url{https://doi.org/#1}}\fi
\ifx \betal  \undefined \def \betal{\textit{et al.}}\fi
\ifx \binstitute  \undefined \def \binstitute#1{#1}\fi
\ifx \binstitutionaled  \undefined \def \binstitutionaled#1{#1}\fi
\ifx \bctitle  \undefined \def \bctitle#1{#1}\fi
\ifx \beditor  \undefined \def \beditor#1{#1}\fi
\ifx \bpublisher  \undefined \def \bpublisher#1{#1}\fi
\ifx \bbtitle  \undefined \def \bbtitle#1{#1}\fi
\ifx \bedition  \undefined \def \bedition#1{#1}\fi
\ifx \bseriesno  \undefined \def \bseriesno#1{#1}\fi
\ifx \blocation  \undefined \def \blocation#1{#1}\fi
\ifx \bsertitle  \undefined \def \bsertitle#1{#1}\fi
\ifx \bsnm \undefined \def \bsnm#1{#1}\fi
\ifx \bsuffix \undefined \def \bsuffix#1{#1}\fi
\ifx \bparticle \undefined \def \bparticle#1{#1}\fi
\ifx \barticle \undefined \def \barticle#1{#1}\fi
\bibcommenthead
\ifx \bconfdate \undefined \def \bconfdate #1{#1}\fi
\ifx \botherref \undefined \def \botherref #1{#1}\fi
\ifx \url \undefined \def \url#1{\textsf{#1}}\fi
\ifx \bchapter \undefined \def \bchapter#1{#1}\fi
\ifx \bbook \undefined \def \bbook#1{#1}\fi
\ifx \bcomment \undefined \def \bcomment#1{#1}\fi
\ifx \oauthor \undefined \def \oauthor#1{#1}\fi
\ifx \citeauthoryear \undefined \def \citeauthoryear#1{#1}\fi
\ifx \endbibitem  \undefined \def \endbibitem {}\fi
\ifx \bconflocation  \undefined \def \bconflocation#1{#1}\fi
\ifx \arxivurl  \undefined \def \arxivurl#1{\textsf{#1}}\fi
\csname PreBibitemsHook\endcsname

%%% 1
\bibitem[\protect\citeauthoryear{Sugon et~al.}{2024}]{SugonBennettMcNamara2024}
\begin{barticle}
\bauthor{\bsnm{Sugon, Jr}, \binits{Q.}},
\bauthor{\bsnm{Bennett}, \binits{C. D.}},
\bauthor{\bsnm{McNamara}, \binits{D. J.}}:
\batitle{Hydrogen Atom as a Nonlinear Oscillator Under Circularly Polarized Light: Epicyclical Electron Orbits}.
\bjtitle{ArXiv preprint arXiv.2410.00056}
%\bvolume{7}(\bissue{3}),
%\bfpage{92}%--\blpage{10807}
(\byear{2024})
\doiurl{10.48550/arXiv.2410.00056}
\end{barticle}
\endbibitem

\bibitem[\protect\citeauthoryear{Sugon~Jr
  et~al.}{2008}]{SugonBragaisMcNamara2008}
\begin{barticle}
\bauthor{\bsnm{Sugon~Jr}, \binits{Q.M.}},
\bauthor{\bsnm{Bragais}, \binits{S.}},
\bauthor{\bsnm{McNamara}, \binits{D.J.}}:
\batitle{Copernicus's epicycles from newton's gravitational force law via
  linear perturbation theory in geometric algebra}.
\bjtitle{arXiv preprint arXiv:0807.2708}
(\byear{2008})
\doiurl{10.48550/arXiv.0807.2708}
\end{barticle}
\endbibitem

%%% 2
\bibitem[\protect\citeauthoryear{Sugon~Jr et~al.}{2026}]{SugonBennettMcNamara2026}
\begin{barticle}
\bauthor{\bsnm{Sugon, Jr}, \binits{Q.}},
\bauthor{\bsnm{Bennett}, \binits{C. D.}},
\bauthor{\bsnm{McNamara}, \binits{D. J.}}:
\batitle{Hydrogen Atom as a Nonlinear Oscillator Under Circularly Polarized Light: Epicyclical Electron Orbits}.
\bjtitle{Hydrogen}
\bvolume{7}(\bissue{3}),
\bfpage{92}%--\blpage{10807}
(\byear{2026})
\doiurl{10.3390/hydrogen7030092}
\end{barticle}
\endbibitem



\end{thebibliography}
%\input sn-article.bbl

\end{document}